# Image Integrity Authentication Scheme Based On Fixed Point Theory

Xu Li[a,d], Xingming Sun[b], Quansheng Liu[c]

[a] College of Information Science and Engineering, Hunan University, Changsha, 410082, China (e-mail: lixu_csust@ 126.com).
[b] College of Computer and Software, Nanjing University of Information Science & Technology, Nanjing, 210044, China (e-mail:sunnudt@163.com).
[c] Laboratoire de Mathématiques de Bretagne Atlantique, Universit´e de Bretagne-Sud, Campus de Tohaninic, BP 573, 56017 Vannes, France (e-mail: quansheng.liu@univ-ubs.fr).
[d] College of Mathematics and Computation Science, Changsha University of Science & Technology, Changsha, 410014, China.

**Abstract:** Based on fixed point theory, this paper proposes a new scheme for image integrity authentication, which is different from Digital Signature and Fragile Watermarking. A realization of the new scheme is given based on Gaussian Convolution and Deconvolution (GCD) functions. For a given image, if it is invariant under a GCD function, we call it GCD fixed point image. An existence theorem of fixed points for GCD functions is proved and an iterative algorithm is presented for finding fixed points. Experiments show that GCD fixed point images perform well in transparence, fragility, security and tampering localization.

## I. INTRODUCTION

Image integrity authentication is used for detecting the inadvertent or malicious modifications to the images transformed over the public channel in order to ensure the communication security. There are two main methods for image integrity authentication: Digital Signature [1] and Fragile

Watermarking [2]. Digital signature, mainly using cryptography techniques, has been playing a very important role in the fields of integrity authentication, identity authentication, etc. Fragile watermarking, a branch of steganography, is now an effective method for integrity authentication and tamper detection.

Compared with fragile watermarking, digital signature has the better performance in whole, and is fairly practical. But in the integrity authentication aspect, we think that fragile watermarking performs better than digital signature in general. In detail, the advantage of digital signature lies in keeping the original image unchanged, high security of key agreement. While the disadvantage lies in additional signature information, being not suitable for tampering localization. The advantage of fragile watermarking lies in transparence, tampering localization and some degree of robustness. While the disadvantage includes modifying the original image slightly and irreversibly in general, complicated agreement of key and watermarking, low security. Till now, people show great enthusiasm in fragile watermarking, and achieve good results [3]-[11]. At the same time, semi-fragile watermarking [12]-[15] has attracted more and more considerations due to its applications in content authentication.

On the view of mathematics, the authentication process of fragile watermarking can be described as follows. Let $I$ be the original image, and $w$ be the watermark, which can be relevant to $I$ or not. For a given secret key $k$, the stego-image can be indicated as $J = H(I, w, k)$, where the function $H(\cdot, \cdot, \cdot)$ represents the embedding process. Let $J'$ be the suspicious image that the receiver received, and $w' = D(J', k)$ be the watermarks extracted from $J'$ using the key $k$, where the function $D(\cdot, \cdot)$ represents the extracting process. If $w = w'$, then we conclude that $J = J'$, which means that the stego-image has not been tampered in transmission.

So far, however, nobody has defined accurate mathematical expressions for $H(\cdot, \cdot, \cdot)$ and $D(\cdot, \cdot)$. We believe that this would restrict further study for fragile watermarking, especially for semi-fragile watermarking. Let $F(\cdot)$ be a filter function and $\bar{J} = F(J)$. If $H(\cdot, \cdot, \cdot)$ or $D(\cdot, \cdot)$ has mathematical expressions, then the semi-fragile problem can be considered as a commutativity problem between functions. For example, if the equation $F(H(I, w, k)) = H(F(I), w, k)$ is true, then the algorithm is an ideal semi-fragile watermarking algorithm for resisting the filter $F(\cdot)$, because $w = D(\bar{J}, k)$ is always true under this condition. While if $D(\cdot, \cdot)$ satisfies $D(F(J), k) = F(D(J, k))$, then we have $w' = F(w)$, and we consider it as a kind of semi-fragile watermarking algorithm too.

The above considerations inspired us to study integrity authentication mathematically. Since digital signature cannot tolerate any image process operation or JPEG compression, and fragile watermarking process has no mathematical expression, it is necessary to find new authentication scheme, which can transform an authentication problem to a problem of function operation.

Based on fixed point theory, this paper proposes a new scheme for image integrity authentication. A realization of the new scheme is given with careful considerations on Gaussian convolution and Gaussian deconvolution. The new scheme is similar to symmetric cryptography, and has better characteristics than fragile watermarking for integrity authentication. The theoretic analysis and simulation results show that the new scheme has great value of development and application, and has good prospects.

The remainder of this paper is organized as follows. In Section Ⅱ, the new scheme of image integrity authentication is given based on fixed point theory. In Section Ⅲ, primary knowledge about Gaussian convolution and Gaussian deconvolution is reviewed, and the GCD function is constructed. In Section Ⅳ, a theorem for the existence of fixed points for GCD function is proved. Based on the existence theorem, an algorithm of image integrity authentication is given in Section Ⅴ, where we also present experimental results to demonstrate the effectiveness of the algorithm in security, transparency, fragility and tempering localization. Section Ⅵ concludes the paper.

## II. IMAGE AUTHENTICATION SCHEME BASED ON FIXED POINT THEORY

For a given mapping $f: D \to D$, if there exists $x \in D$ such that $f(x) = x$, then $x$ is called a fixed point of the mapping $f$ in the space $D$. We consider the case where $x$ is an image, $f$ an operation on images, and $D$ a space of images. Usually the set of fixed points of $f$ is sparse in $D$, so that almost any change will make a fixed point image to be a non-fixed point image, since a fixed point image has very little chance to be changed into another fixed point due to the sparsity of the fixed points. This means that a fixed point image has the fragility property. Therefore fixed point images can be used as stego-image during the authentication process.

The mathematical model of the image integrity authentication scheme based on fixed point theory is proposed as follows. We consider grayscale images in this paper. An 8-bits image $I$ of size $M \times N$ can be indicated as $I \in \mathbf{Z}_{256}^{M \times N}$, where $\mathbf{Z}_{256} = \{0, 1, \cdots, 255\}$. Let $I \in \mathbf{Z}_{256}^{M \times N}$ be a given image, $K$ be the key space. For all $k \in K$, we define $f_k(\cdot): \mathbf{Z}_{256}^{M \times N} \to \mathbf{Z}_{256}^{M \times N}$. Then the function $f_k(\cdot)$ can be used for image integrity authentication if it satisfies the following conditions:

1) the function $f_k(\cdot)$ has sparse fixed points in space $\mathbf{Z}_{256}^{M \times N}$;

2) after finite iterations, the image $I$ can reach a fixed point $J$, i.e. $f_k(J) = J$;

3) the original image $I$ and the fixed point image $J$ should not be distinguished visually.

The security of the function $f_k(\cdot)$ entirely relies on the secret key $k$, and if the key space $K$ is large enough, then the scheme is safe even when the algorithm is public.

The first condition emphasizes the existence of fixed point, which is the foundation of the scheme. While if $f_k(\cdot)$ has no fixed point, then the scheme will be impractical. The fragility of the fixed point image $J$ based on the sparsity of the fixed points, while if the fixed points are dense in space $\mathbf{Z}_{256}^{M \times N}$, the scheme would be vulnerable to attacks and there would be missed alarms. The second condition emphasizes the practicality, which means that any given image can converge to a fixed point image shortly after several iterations, so that the scheme can efficiently treat a lot of images. The third condition emphasizes the transparency, in other words, the fixed points of function $f_k(\cdot)$ should not be sparse excessively. Otherwise, it would lead to an large difference between $I$ and $J$, and make the integrity authentication lose its meaning.

Let's now consider the process of authentication. Firstly, the sender and the receiver establish the private key $k$. Secondly, the sender sends $J$ instead of $I$ to the receiver through public channel, where $J$ is the corresponding fixed point image generated from $I$ by the function $f_k(\cdot)$. Thirdly, for the suspicious received image $J'$ (which may be the fixed point image $J$ or not, because $J$ may be tampered with in transmission), the receiver calculates $f_k(J')$ and compares the result to $J'$, and then judges the integrity of the fixed point image $J$. If $J'$ is a fixed point of $f_k(\cdot)$, then we conclude that $J = J'$. When $J'$ is not a fixed point, $f_k(J')$ will change $J'$ and make it approximate to the nearest fixed point of $f_k(\cdot)$, and if $f_k(\cdot)$ is well-chosen, the set $\{J'(s,t): f_k(J') \neq J'\}$ of pixels where the two images $J'$ and $f_k(J')$ are different can include some of the distortion status of image $J$. So we say that the authentication scheme can be used for tempering localization for the well-chosen function $f_k(\cdot)$.

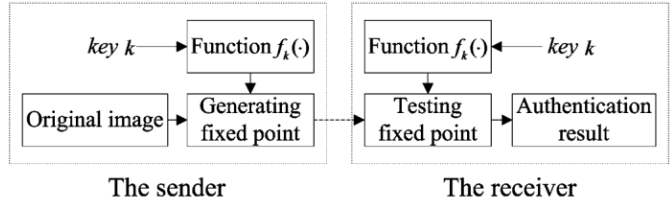

Fig. 1. Framework of integrity authentication using fixed point images.

The framework of the authentication scheme based on fixed point theory is presented in Fig. 1. This scheme is simpler than fragile watermarking and digital signature, since it does not embed watermarks into the original image or produce extra signature information from the original image. In this authentication scheme, the most important thing is to find out a suitable function $f_k(\cdot)$. In the following section, a function meeting the above three conditions is constructed with Gaussian convolution and Gaussian deconvolution.

## III. PRELIMINARY KNOWLEDGE

Gaussian convolution [16] can effectively filter out Gaussian noise in images corrupted by Gaussian noise, and is mainly used in image restoration and edge detection. Gaussian deconvolution is used for removing low pass filter in images, and is also a kind of image restoration algorithm. As inverse problem, Gaussian deconvolution is usually more difficult in computation than Gaussian convolution, and is still a challenging problem. In this paper, as a mature algorithm for Gaussian convolution and Gaussian deconvolution, Fourier transform is used in computation. It translates between convolution and multiplication of functions, and makes the computation of deconvolution as simple as that of convolution in certain cases.

In one dimension space, for an image $I$ and a Gaussian kernel

$$g_\sigma(x) = \frac{1}{\sqrt{2\pi}\sigma} e^{-\frac{x^2}{2\sigma^2}}, \quad \sigma > 0, \tag{1}$$

the Gaussian convolution is defined as

$$G_\sigma(I)(x) = I * g_\sigma(x) = \int_{-\infty}^{+\infty} I(x-t)g_\sigma(t)dt. \qquad (2)$$

Gaussian convolution is perhaps the most often used image operator in image processing, but it smoothes an image and blurs it at the same time.

Gaussian deconvolution is to restore the original image $I$ when $g_\sigma$ and $G_\sigma(I)$ are given. It is more difficult in computation than Gaussian convolution, and has usually no unique solution, so that Gaussian convolution and Gaussian deconvolution cannot be inverse operations in general.

In practical computation of convolution (resp. deconvolution), Fourier transform is an efficient tool, which translates convolution (resp. multiplication) problem in spatial domain to multiplication (resp. convolution) problem in frequency domain, via the famous convolution theorem:

$$F(I * h) = F(I) \cdot F(h), \qquad (3)$$

$$F(I \cdot h) = F(I) * F(h). \qquad (4)$$

Where $F(\cdot)$ denotes the Fourier transform and $h(x)$ represents an image function. In addition, we assume $F^{-1}(\cdot)$ denotes the inverse Fourier transform.

Thus the $G_\sigma(\cdot)$ and the $G_\sigma^{-1}(\cdot)$ (the corresponding Gaussian deconvolution) can be formulated by Fourier transform as follows (using the convolution theorem):

$$G_\sigma(I) = F^{-1}[F(I) \cdot F(g_\sigma)], \qquad (5)$$

$$G_\sigma^{-1}(I) = F^{-1}[F(I)/F(g_\sigma)]. \qquad (6)$$

Note that when the denominator $F(g_\sigma)$ has no zero points, the deconvolution $G_\sigma^{-1}(\cdot)$ is well defined. In this paper we will always suppose this whenever $G_\sigma^{-1}(\cdot)$ is concerned. The convolution and deconvolution are inverse operations, namely $I = G_\sigma^{-1}[G_\sigma(I)] = G_\sigma[G_\sigma^{-1}(I)]$. Throughout the paper, we take $\sigma \leq 0.4246$ (this constant is explained in Section IV), so that $F(g_\sigma)$ has no zero points, and the Gaussian deconvolution is feasible and practical.

The properties of Fourier transform show that $G_\sigma(\cdot)$ and $G_\sigma^{-1}(\cdot)$ are both linear operators:

$$G_\sigma(I + h) = G_\sigma(I) + G_\sigma(h), \qquad (7)$$

$$G_\sigma^{-1}(I + h) = G_\sigma^{-1}(I) + G_\sigma^{-1}(h). \qquad (8)$$

Because the image pixel values are integers, rounding and truncation must be executed to the result of convolution or deconvolution. In order to express clearly the computation process, we introduce an integer function $R(x)$: let $\mathbf{Z}$ be the integer space, for all $n \in \mathbf{Z}$ and $x \in [n, n+1]$, we define

$$R(x) = \begin{cases} n & \text{if } x - n < 0.5 \\ n+1 & \text{if } x - n \geq 0.5 \end{cases}. \tag{9}$$

For a given image $I$ and $\sigma > 0$, we define the Gaussian Convolution and Deconvolution (GCD) function as follows:

$$\boxed{f_\sigma(I) = R\left[G_\sigma^{-1}\left[R(G_\sigma(I))\right]\right]}. \tag{10}$$

The detailed computation process in frequency domain can be described as

$$f_\sigma(I) = R\left\{F^{-1}\left[F\left(R\{F^{-1}[F(I) \cdot F(g_\sigma)]\}\right)\middle/ F(g_\sigma)\right]\right\}. \tag{11}$$

Generally, $I$ and $f_\sigma(I)$ are not identical, and the difference between them increases with the increase of $\sigma$. Sometimes, $f_\sigma(I)$ may jump out of the given image space of $I$. As notation, we still use the above signs for two dimension problems.

## IV. Fixed Points of GCD Functions

Firstly we give the mathematical signs used in what follows. Let $\mathbf{R}^{M \times N}$ and $\mathbf{Z}^{M \times N}$ be respectively the $M \times N$ dimensional real space and integer space. Set $\mathbf{Z}_{256} = \{0, 1, \cdots, 255\}$. Then for any 8-bits grayscale image $I$ of size $M \times N$, we have $I \in \mathbf{Z}_{256}^{M \times N} \subset \mathbf{Z}^{M \times N} \subset \mathbf{R}^{M \times N}$. As further notation, we write $I = a\, (a \in \mathbf{R})$ if each matrix element of $I$ is equal to $a$, We denote by $|I|$ the matrix whose elements are the absolute values of the corresponding elements of $I$, and we write $\min(I)$ (resp. $\max(I)$) for the minimum (resp. maximum) value of the elements of $I$.

**Definition.** For $\sigma > 0$ and $I \in \mathbf{Z}^{M \times N}$, if $I = f_\sigma(I)$, then $I$ is called a GCD fixed point of $f_\sigma(\cdot)$.

**Lemma 1.** (a) For $0 < \sigma \leq 1/2\sqrt{\ln 2}$ and $I \in \mathbf{R}^{M \times N}$, if $I$ is bounded, i.e. $I(s,t) \in [a,b]$, where $(s,t)$ is the spatial coordinates, then we have

$$a \leq I(s,t) - 4e^{-\frac{1}{2\sigma^2}}\left(I(s,t) - a\right) \leq \left[G_\sigma(I)\right](s,t)$$

$$\leq I(s,t)+4e^{-\frac{1}{2\sigma^2}}\left(b-I(s,t)\right)\leq b. \tag{12}$$

(b) Suppose that $I \in \mathbf{R}^{M \times N}$ is random with uniform distribution $U[-0.5, 0.5]$, we have $G_\sigma(I) \in [-0.5, 0.5]$ and $P\{G_\sigma(I) \in (-0.5, 0.5)\} = 1$.

**Proof.** (a) By the definition of Gaussian convolution, we have $a \leq [G_\sigma(I)](s,t) \leq b$ that holds for all $\sigma > 0$. When $0 < \sigma \leq 1/2\sqrt{\ln 2}$, by direct calculation, we see that

$$[G_\sigma(I)](s,t) = \frac{1}{h}\sum_{i=1}^{M}\sum_{j=1}^{N} I(i,j) \frac{1}{2\pi\sigma^2} e^{-\frac{(i-s)^2+(j-t)^2}{2\sigma^2}}$$

$$\geq I(s,t) - \frac{I(s,t)-a}{2\pi\sigma^2 h} \sum_{\substack{i=1 \\ i \neq s}}^{M}\sum_{\substack{j=1 \\ j \neq t}}^{N} e^{-\frac{(i-s)^2+(j-t)^2}{2\sigma^2}}$$

$$\geq I(s,t) - \frac{I(s,t)-a}{2\pi\sigma^2 h} \frac{4e^{-\frac{1}{2\sigma^2}}}{1-e^{-\frac{1}{2\sigma^2}}}, \tag{13}$$

where $h > 0$ is the normalization coefficient of the convolution template, which satisfies

$$h = \sum_{i=1}^{M}\sum_{j=1}^{N} \frac{1}{2\pi\sigma^2} e^{-\frac{(i-s)^2+(j-t)^2}{2\sigma^2}} \geq \frac{1}{2\pi\sigma^2}. \tag{14}$$

So we have

$$[G_\sigma(I)](s,t) \geq I(s,t) - 4e^{-\frac{1}{2\sigma^2}}\left(I(s,t)-a\right). \tag{15}$$

Similarly, we get

$$[G_\sigma(I)](s,t) \leq I(s,t) + 4e^{-\frac{1}{2\sigma^2}}\left(b-I(s,t)\right). \tag{16}$$

(b) By the preceding discussion, this part can be proved easily.

**Remark 1.** If $I$ takes random values in $\mathbf{R}^{M \times N}$ according to the uniform distribution on some interval $[a,b]$ with length of 1, we assume naturally that $\{I\} = R(I) - I$ follows the uniform distribution $U[-0.5, 0.5]$.

**Lemma 2.** For $\sigma > 0$ and $I \in \mathbf{Z}^{M \times N}$, $J = R\left[G_\sigma^{-1}(I)\right]$ is a GCD fixed point with probability 1.

**Proof.** By Remark 1, we have $J = G_\sigma^{-1}(I) + \{J\}$, where $\{J\}$ follows the uniform distribution $U[-0.5, 0.5]$. Hence

$$R\left[G_\sigma^{-1}\left[R(G_\sigma(J))\right]\right] = R\left[G_\sigma^{-1}\left[R\left(G_\sigma\left(G_\sigma^{-1}(I) + \{J\}\right)\right)\right]\right]$$

$$= R\left[G_\sigma^{-1}\left[R\left(G_\sigma\left(G_\sigma^{-1}(I)\right) + G_\sigma(\{J\})\right)\right]\right]$$

$$= R\left[G_\sigma^{-1}\left[R\left(I + G_\sigma(\{J\})\right)\right]\right]. \tag{17}$$

Therefore by Lemma 1(b), with probability 1, we have

$$R\left[G_\sigma^{-1}\left[R(G_\sigma(J))\right]\right] = R\left[G_\sigma^{-1}(I)\right] = J. \tag{18}$$

**Lemma 3.** For $\sigma > 0$ and $I \in \mathbf{Z}^{M \times N}$, $J = R\left[G_\sigma^{-1}\left[R(G_\sigma(I))\right]\right]$ is a GCD fixed point with probability 1. Furthermore, $G_\sigma(I - J) \in [-1, 1]$.

**Proof.** The first conclusion follows directly from Lemma 2. For the second conclusion, we notice that

$$G_\sigma(I - J) = G_\sigma\left(I - R\left[G_\sigma^{-1}\left[R(G_\sigma(I))\right]\right]\right)$$

$$= G_\sigma\left(I - G_\sigma^{-1}\left[R(G_\sigma(I))\right] + \{J\}\right)$$

$$= G_\sigma(I) - R\left[G_\sigma(I)\right] + G_\sigma(\{J\})$$

$$= \{G_\sigma(I)\} + G_\sigma(\{J\}). \tag{19}$$

By Lemma 1(b), we have $G_\sigma(I - J) \in [-1, 1]$.

**Theorem 1.** For all $I \in \mathbf{Z}^{M \times N}$, if $0 < \sigma \leq 0.4246$, then there exists $J \in \mathbf{Z}^{M \times N}$, such that:

(a) $J$ is a GCD fixed point with probability 1;

(b) $|I - J| \leq 1$.

**Proof.** Let $J = R\left[G_\sigma^{-1}\left[R(G_\sigma(I))\right]\right]$. By Lemma 3, Part (a) holds evidently. For the proof of Part (b), note that both $I$ and $J$ are in $\mathbf{Z}^{M \times N}$, so that $-k \leq I - J \leq k$, where $k = \max(|I - J|)$. Choose $(s, t)$ such that $|(I - J)(s, t)| = k$. Assume that $(I - J)(s, t) = k$ (the case where $(I - J)(s, t) = -k$ can be considered in a similar way). Then by Lemma 1(a), we have

$$k - 8ke^{-\frac{1}{2\sigma^2}} \leq [G_\sigma(I-J)](s,t) \leq k. \tag{20}$$

Since $G_\sigma(I-J) \in [-1,1]$ by Lemma 3, the left side of the above inequality satisfies $k - 8ke^{-\frac{1}{2\sigma^2}} \leq 1$, which is equivalent to $\sigma \geq \sqrt{(2[\ln(8k) - \ln(k-1)])^{-1}}$. When $k=2$ it reads $\sigma \geq 1/2\sqrt{2\ln 2} = 0.42466...$. Therefore $\sigma \geq 0.42466...$ when $k \geq 2$. It follows that when $\sigma \leq 0.4246$, we have $k \leq 1$, so that $|I-J| \leq 1$.

**Remark 2.** By Theorem 1, if an image $I \in \mathbf{Z}_{256}^{M \times N}$ has some pixels taking values in $\{0, 255\}$, then $J = R\left[G_\sigma^{-1}\left[R(G_\sigma(I))\right]\right]$ may jump out of the image space $\mathbf{Z}_{256}^{M \times N}$, because these pixel values may be changed into $-1$ or $256$. But if all the pixels of $I$ take values in $\{1, \cdots, 254\}$, and if $\sigma \leq 0.4246$, then $J$ is a GCD fixed point image with probability 1, and $|I-J| \leq 1$. This means that there are about $84^{M \times N}$ different GCD fixed points in $\mathbf{Z}_{256}^{M \times N}$ when $\sigma \leq 0.4246$, and the sparsity is established. If someone tries to find out a fixed point randomly, his success probability is about $1/3^{M \times N}$, so that it is almost impossible for him to succeed. From the discussion above, we see that when $\sigma \leq 0.4246$, the GCD function satisfies the three conditions of Section II, and we can use it for image integrity authentication.

**Theorem 2.** For $\sigma > 0$ and $I \in \mathbf{Z}_{256}^{M \times N}$, suppose that $T_\theta(I)$ represents the rotation of the image $I$ by $\theta$ degrees, where $\theta \in \{k\pi/2 : k \in \mathbf{Z}\}$. If $G_\sigma^{-1}(\cdot)$ is well defined, then $T_\theta(\cdot)$ and the GCD function $f_\sigma(\cdot)$ are commutative, i.e.

$$T_\theta[f_\sigma(I)] = f_\sigma[T_\theta(I)]. \tag{21}$$

**Proof.** By the rotation character of Fourier transform, rotating $I$ by $\theta$ degrees leads to the same rotation of $F(I)$ (the Fourier transform of $I$), which means $T_\theta(I) = F^{-1}\{T_\theta[F(I)]\}$. Since $g_\sigma$ (the Gaussian kernel) is central symmetric and axial symmetric, we have $T_\theta(g_\sigma) = g_\sigma$. So, for the Gaussian convolution $G_\sigma(\cdot)$, by the convolution theorem, we have

$$T_\theta[G_\sigma(I)] = T_\theta\{F^{-1}[F(I) \cdot F(g_\sigma)]\}$$

$$\begin{aligned}
&= F^{-1}\left[T_\theta\left(F(I)\cdot F(g_\sigma)\right)\right] \\
&= F^{-1}\left[F\left(T_\theta(I)\right)\cdot F\left(T_\theta(g_\sigma)\right)\right] \\
&= T_\theta(I)*T(g_\sigma) = T_\theta(I)*g_\sigma \\
&= G_\sigma[T_\theta(I)]. \quad\quad\quad\quad\quad\quad\quad\quad\quad\quad\quad\quad\quad\quad\quad\quad (22)
\end{aligned}$$

Similarly, we have $T_\theta\left[G_\sigma^{-1}(I)\right] = G_\sigma^{-1}[T_\theta(I)]$ for the Gaussian deconvolution $G_\sigma^{-1}(\cdot)$. Since $T_\theta[R(I)] = R[T_\theta(I)]$ holds obviously for $R(\cdot)$, we get $T_\theta[f_\sigma(I)] = f_\sigma[T_\theta(I)]$ directly.

**Theorem 3.** For all $\sigma > 0$ and any given image $I$, if $G_\sigma^{-1}(\cdot)$ is well defined, then the transpose operation and the GCD function $f_\sigma(\cdot)$ are commutative, i.e. $[f_\sigma(I)]^T = f_\sigma(I^T)$.

**Proof.** The proof is just like that of Theorem 2, using the fact that $F(I^T) = [F(I)]^T$ and $(g_\sigma)^T = g_\sigma$, which can be verified easily.

**Remark 3.** By Theorems 2 and 3, we see that GCD fixed point images can resist some geometric attacks, such as transpose attack, some rotation attacks and flipping attacks. As we expected in Section Ⅰ, the resistance processes can be abstracted into commutativity problem between functions.

## V. Algorithm for Image Integrity Authentication Based on GCD function

In this section, based on the previous discussions, an iteration algorithm for approaching fixed point images of GCD function is given. In order to reduce the computation cost and to enhance the security of the symmetric key, the Gaussian filter in the frequency domain is introduced, and a secure key generating scheme is designed based on the standard deviation $\sigma'$ of the filter function.

Notice that the Fourier transform of $g_\sigma$ is also a Gaussian function. Therefore, we use the Gaussian filter in the frequency domain instead of the Gaussian convolution kernel in the spatial domain to calculate the convolution and deconvolution. We denote the Gaussian filter as $g_{\sigma'}$, and construct it directly in the frequency domain. For an image of size $M \times N$, Simulations show that $\sigma'$ has an inverse relationship with $\sigma$, and has positive relationships with $M$ and $N$.

With the Gaussian filter $g_{\sigma'}$ and the center shift Discrete Fourier Transform (DFT), the Gaussian

convolution and the Gaussian deconvolution in (5) and (6) can be expressed as:

$$G_{\sigma'}(I) = F^{-1}[F(I) \cdot g_{\sigma'}], \quad (23)$$

$$G_{\sigma'}^{-1}(I) = F^{-1}[F(I)/g_{\sigma'}]. \quad (24)$$

Then, the practical computation process for the GCD function in (10) can be described as

$$f_{\sigma'}(I) = R\{F^{-1}[F(R\{F^{-1}[F(I) \cdot g_{\sigma'}]\})/g_{\sigma'}]\}. \quad (25)$$

Considering the fixed point problem with the new GCD function $f_{\sigma'}(\cdot)$, we can get the conclusions similar to Theorems 1, 2 and 3.

## A. Security

The authentication scheme depends on the security of the secret key completely, so the key space should be necessarily large enough. Disappointingly, if $\sigma$ (in the spatial domain) or $\sigma'$ (in the frequency domain) is used simply as the secret key, the corresponding key space cannot meet the security requirements. For example, given a grayscale image of size $512 \times 512$ (in this situation, $\sigma = 0.4246$ corresponding to $\sigma' = 361$ approximately), for the parameter $\sigma'$ in frequency domain, simulations show that the function $f_{\sigma'}(\cdot)$ has too many fixed points when $\sigma' > 1000$, which results in poor fragility. While when $\sigma' < 250$, the function $f_{\sigma'}(\cdot)$ will have too few fixed points, which results in a lot of iterative calculations. In addition, when $\sigma'$ is accurate to four decimal places, two adjacent parameters may have some of the same fixed points. In conclusion, for an image of size $512 \times 512$, the standard deviation $\sigma'$ should take values between $250$ and $1000$, and be accurate to three decimal places. Under these conditions, the size of the key space is about $2^{19}$, which is too small to prevent exhaustive search.

A security, efficient key generating scheme based on the parameter $\sigma'$ is proposed as follows. Firstly, for an image of size $M \times N$, construct a Gaussian filter with the parameter $\sigma' = r \times (M + N)$, where $r \in (0,1)$; secondly, properly choose some points in the filter; thirdly, properly modify the values of these points. Then we get a new filter which should be called artifact Gaussian filter, and the new secret key is generated by adding the modification information to the parameter $r$. For example, given a grayscale image of size $512 \times 512$, let $r = 0.500000$, then we have $\sigma' = 512.000$.

For the filter constructed with $\sigma'$, if we change the values of the points $(1,20)$ and $(30,100)$ to $0.6$ and $0.5$ separately, then we can write the secret key as follows:

**0.500000 1 20 0.6 30 100 0.5**.

The size of the key space is about $2^{19} \times 512^4 \times 10^2 \approx 2^{61}$. If $n$ points in the Gaussian filter are chosen and the corresponding values are modified, the size of the key space is about $2^{19} \times M^n \times N^n \times 10^n$. Obviously, the key space is large enough as long as $(M, N, n)$ is not too small.

Some principles should be followed when we modify the Gaussian filter. ***On the premise of ensuring the security of the key generating scheme, the filter should be modified as less as possible, and at the same time, the chosen points should be far away from the center of the filter.*** Because the positions around the center correspond to the low frequency components of the images, so too many modifications of the filter may influence the iterative calculation and the quality of the fixed point images.

*B. Algorithm for image integrity authentication*

Based on Theorem1 and Remark 2, we propose the following algorithm for image integrity authentication.

**Algorithm**

The sender and the receiver agreed on a secret key $k$ in advance.

**For the Sender:**

**Step 1.** For any image $I$, Generate Gaussian (or artifact Gaussian) filter using the secret key $k$;

**Step 2.** Calculate $J = f_k(I)$ for the original image $I \in \mathbf{Z}_{256}^{M \times N}$;

**Step 3.** Judge the range of $J$, if $J \in \mathbf{Z}_{256}^{M \times N}$, continue to step 4; if not, adjust $I$ according to $J$ (adjusting method is as follows, where $a=1$, and if the program falls into an infinite loop, then let $a = a+1$), and pass to step 2;

$$I(s,t) = \begin{cases} a & \text{if } J(s,t) < 0 \\ 255-a & \text{if } J(s,t) > 255 \end{cases}. \tag{26}$$

**Step 4.** Compare $J$ and $I$ for equality; if $J = I$, then send $J$ to the receiver over a public channel; if not, let $I = J$ and pass to step 2.

**For the Receiver:**

**Step 1.** For the received image $J'$, Generate Gaussian (or artifact Gaussian) filter using the secret key $k$;

**Step 2.** Calculate $J'' = f_k(J')$ for the suspicious image $J'$. If $J'' = J'$, then $J$ has not been attacked; if not, then $J$ has been tampered, and the tampered points lie at the points where $J'' \neq J'$.

Let us give some comments on the algorithm. For the sender, steps 2-4 are used for iterative computations and image adjustments, and when the program falls into an infinite loop, the image adjustment strength should be increased. For the receiver, the verification of the received image $J'$ is very simple, because only one GCD function calculation is necessary, and then the integrity can be judged by whether $J'$ is a fixed point of the GCD function. Furthermore, the tampered areas can be localized because $J'$ will approximate the nearest fixed point after the GCD calculation, and the discussion in Remark 2 has also indicated that $J'$ is impossible to be another fixed point different from $J$.

*C. Transparence*

The transparency of fixed point images is essentially determined by $\sigma'$ or $\sigma$. The following experiments show that the transparence increases as the parameter $\sigma'$ (or $\sigma$) increases (decreases). The Peak Signal to Noise Ratio (PSNR) is used to evaluate the quality of images. For the Gaussian filters generated from different $\sigma'$, we obtain the secret keys by simply changing the values of the points (1, 2) and (3, 1) into 0.2 (by the discussion in subsection A, at least 6 points are necessary for security; we choose two points here because other characters of the fixed point images are mainly influenced by $\sigma'$). Image databases FREEFOTO [17] (10408 grayscale images with sizes of 400 ×600 or 600×400) are used for testing.

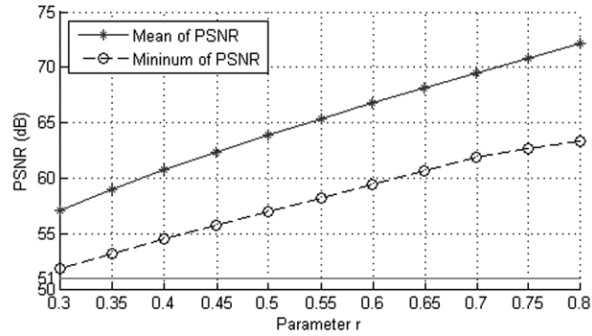

**Fig. 2.** Transparence test for FREEFOTO.

Compared with the results of the fragile (or semi-fragile) watermarking methods [8]-[16], our experiment results (Fig. 2) show that the GCD fixed point images perform well in transparence. The minimal PSNR value of the fixed point images is greater than 51 dB; the absolute difference between the original images and their corresponding fixed point image is less than or equal to 2. In addition, the PSNR value can be adjusted freely in a continuous interval, while in other papers the PSNR can only takes several discrete values.

A statistical result is worth to be mentioned here. In the image database FREEFOTO, when $r$ (or $\sigma'$) takes large values, there are few images which are lossless GCD fixed point images, in other words, these images can be used for integrity authentication without modification. This phenomenon gives us a warning that some attackers may use black box method to cheat the receiver if $r$ is too large. So, the parameter $r$ must be selected carefully.

*D. Fragility*

Contrary to the transparence, the fragility increases as the parameter $\sigma'$ (or $\sigma$) decreases (increases). Select an image of size $400 \times 600$ from FREEFOTO, let the secret key be "0.**500000** **1** **2** **0.5** **3** **1** **0.5** **10** **10** **0.6**". Fig. 3 shows the results of the fragility test according to some general attacks, such as image processing and JPEG compression etc. Binary images are used to illustrate the authentication result, and the suspicious points are revealed by the white spots.

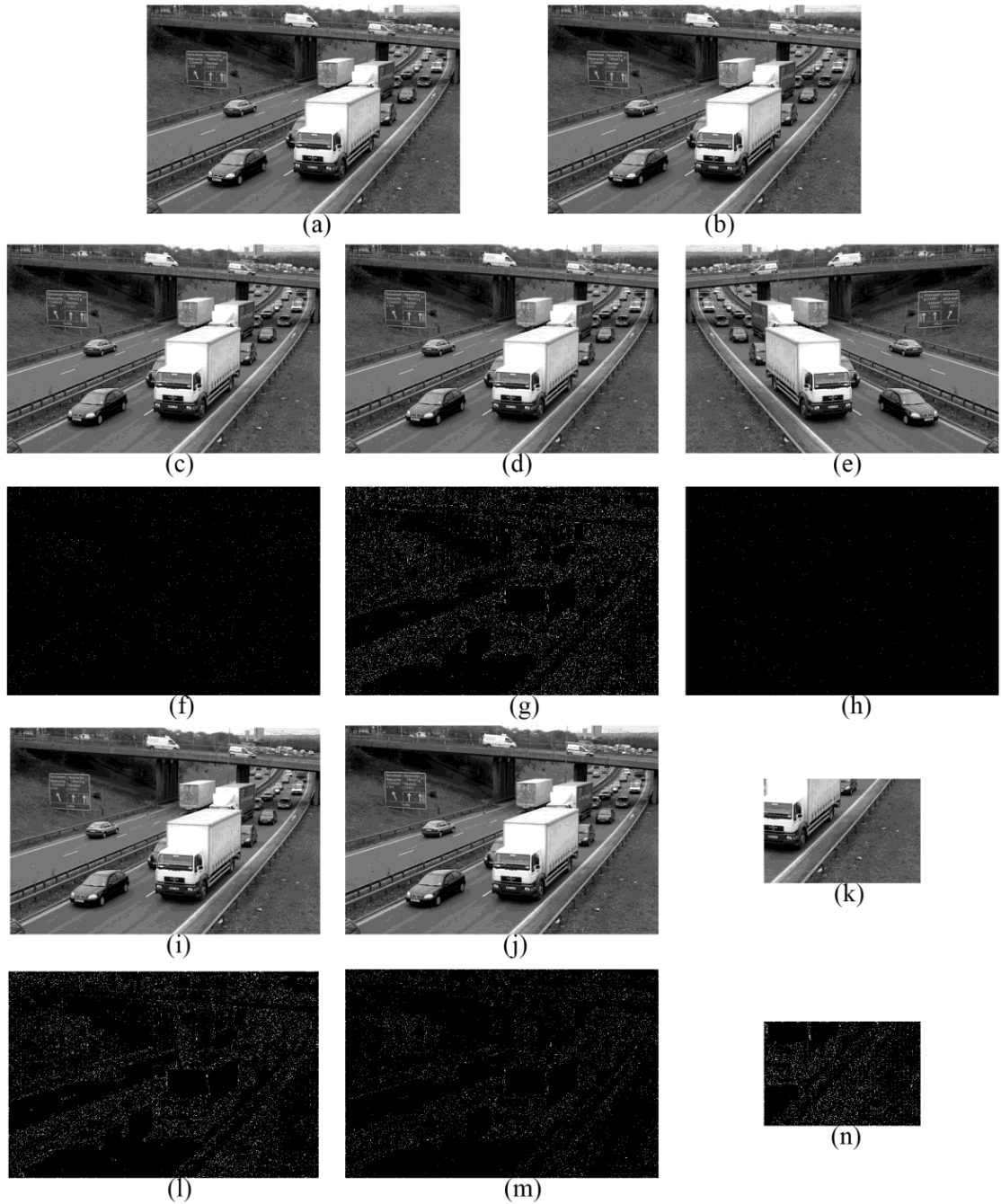

**Fig. 3.** Test for fragility. (a) is the original image of size 400×600; (b) is the fixed point image (PSNR=61.4965 dB); (c) is the Rewriting attack with the known original image ($\sigma' = 501.000$, PSNR=61.5380 dB); (d) is the JPEG compression ( ratio 0.95); (e) is the Flipping attack; (i) is the Scaling attack ( ratio 0.99); (j) is the Histogram enhancement attack; (k) is the Cropping attack; (f), (g), (h), (l), (m)and (n) are the authentication results.

**Remark 4.** Since the artifact Gaussian filter $g_{\sigma'}$ is not central symmetric and axial symmetric, by Theorems 2 and 3, the fixed point images cannot resist the flipping attack in Fig. 3(e). Because the

Gaussian filter is modified slightly, the attack response (Fig. 3(h)) of the fixed point image is slight too. While, if the well-chosen key can generate the artifact Gaussian filter $g_{\sigma'}$ to be central symmetric and axial symmetric, the corresponding fixed point images can resist some geometric attacks.

*E. Tampering localization*

Select another image of size 400×600 from FREEFOTO, and take the secret key as the previous one. Fig. 4 shows the experiment results for tampering localization.

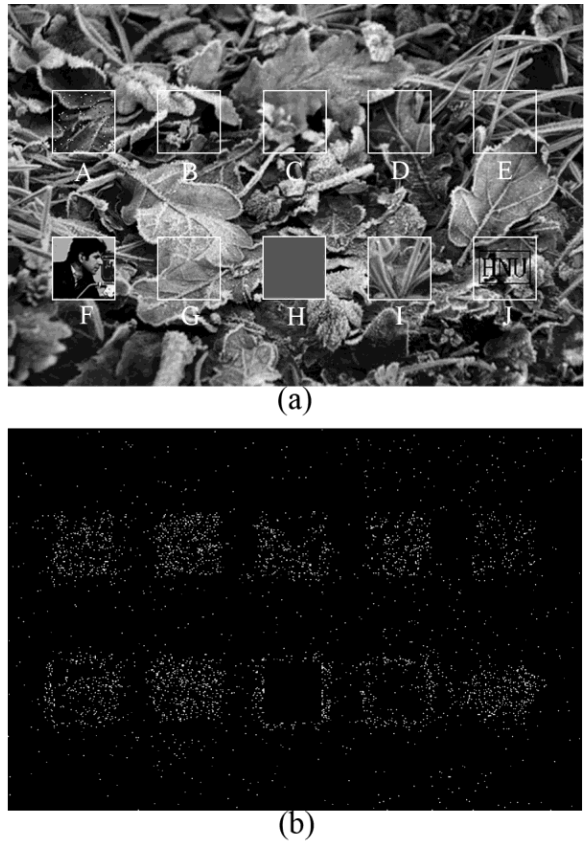

**Fig. 4.** Test for tampering localization. The PSNR value of the fixed point image is equal to 59.7802 dB; the tampered areas are marked with white rectangles. (a) shows some common attacks, where A is the local Salt & Pepper noise attack; B is the local Gaussian noise attack; C is the local Median filter attack; D is the local Gaussian filter attack; E is the local Histogram enhancement attack; F is the Copy attack from an another image; G is the Copy attack from itself; H is the Cover attack with a constant; I is the Collage attack; J is adding illegal Logo; (b) shows the authentication results.

The experiment results show that most of localization errors happen near the edges of the tampered areas. At the same time, there are few, scattered and non-tampered points which may be wrongly marked as suspicious points. As discussed in Section Ⅱ, if a fixed point image is tampered, then after one calculation with the GCD function, the tampered image will approach the nearest fixed point. We cannot ensure that these two fixed point images are the same one, but we see that the GCD function guarantees that each of the pixels of the tampered image is mainly influenced by its neighbors during the calculation. So, we say that most of the wrong localizations are not far away from the tampered areas, and few scattered localization errors happen inevitably.

For fixed point images, the ability of tampering localization is affected by two factors. One is the parameter $\sigma'$ and the other is the modification to the Gaussian filter. In fact, the localization performs better with smaller $\sigma'$ and minor modifications.

Fixed point images can identify the Collage Attack. In fig. 4, when the fixed point image is attacked by the Collage Attack, the authentication result shows a hollow area, which is a special mark. At the same time, the authentication result of the Cover attack shows to be a hollow area too. The reason is that a constant image can be considered as a fixed point image according to all secret keys, so the Cover attack with a constant can be considered as a special Collage attack.

## VI. CONCLUSION

We have proposed a new image integrity authentication scheme based on fixed point theory. In the proposed scheme, the key issue is to find an appropriate function $f_k(\cdot)$ for calculation. We have given a realization based on the GCD function, which is constructed with Gaussian convolution and Gaussian deconvolution. While we believe that there exist more suitable functions waiting to be discovered for this scheme.

Since fragile watermarking is divided into total-fragile watermarking and semi-fragile watermarking, a nature thought is that fixed point image can be divided into total-fragile fixed point image and semi-fragile fixed point image. By Theorems 2, 3 and Remark 4, when the convolution kernel is central symmetric and axial symmetric, the GCD fixed point images are semi-fragile fixed point images. Otherwise, they are total-fragile fixed point images. We hope the discussions about the semi-fragile problem and the problem of commutative functions can bring improvements for the study of Content Authentication.

ACKNOWLEDGEMENT

This work is supported by the NSFC (61232016, 61173141, 61173142, 61173136, 61103215, 61070196, 61070195, and 61073191), National Basic Research Program 973 (2011CB311808), 2011GK2009, GYHY201206033, 201301030, 2013DFG12860 and PAPD fund.